# Necking of epithelial tissues with cellular topological transition

Yuan He[1,2], Shi-Lei Xue[2,#]

**1** School of Materials Science and Engineering, Zhejiang University, Hangzhou, Zhejiang 310027, China, **2** Department of Materials Science and Engineering, School of Engineering, Westlake University, Hangzhou, Zhejiang 310030, China
[#] Corresponding author: xueshilei@westlake.edu.cn.

**Abstract**

As the cover of embryos and adult organisms, epithelial tissues are subjected to substantial mechanical forces in tissue morphogenesis. However, the finite deformation behaviors of epithelial tissues remain largely unexplored. This study combines discrete vertex simulations with a multiscale constitutive model to investigate the necking behavior of epithelial tissues. In the multiscale model, the shape changes and topological transitions of single cells are mapped to the elastic and inelastic tissue deformations via a mean-field formulation. Our results show that the necking bifurcation of a stretched tissue arises from cellular topological transitions. The bifurcation condition and the steady state of necking propagation are predicted from the constitutive model and validated by vertex simulations. Furthermore, we find that topological defects in disordered tissues facilitate necking bifurcation but impede its propagation. These defects also induce the necked region to collapse into a thin thread, as observed in real tissues. Together, our work provides valuable insights into the deformation behaviors of epithelial tissues.

## Author Summary

Epithelial tissues form thin sheets that protect organs and play key roles in development and disease. During morphogenesis and mechanical loading, these tissues often undergo large deformations, including narrowing, thinning, and eventual rupture. Although experiments have observed striking behaviors such as tissue necking, the cellular mechanisms that drive these large-scale mechanical instabilities remain unclear.

Here, we combine a computational vertex model with a multiscale theoretical framework to investigate necking in epithelial tissues under stretching. Our approach explicitly links tissue-scale deformation to cell-scale mechanical events. We show that reversible changes in cell shape produce elastic tissue deformation, whereas irreversible cell rearrangements, known as topological transitions, generate inelastic deformation. Importantly, we find that tissue necking is triggered by these cellular rearrangements rather than by elastic deformation alone.

Our model predicts when necking initiates, how it propagates through the tissue, and how stress relaxes during deformation, and these predictions are quantitatively validated by cell-based simulations. We further demonstrate that topological defects in disordered tissues strongly influence where necking begins and whether it propagates. Together, our results provide a mechanistic link between cell rearrangements and large-scale mechanical instabilities in epithelial tissues, offering insights into tissue morphogenesis and mechanical failure.

# 1. Introduction

Epithelial tissues are one-cell-thick continuous sheets that cover the surfaces of embryos and adult organisms. Recently, mechanics of epithelial tissues have been found to be important for tissue morphogenesis. Subjected to various mechanical forces, epithelial tissues can undergo remarkably large-scale deformations and endow organisms with diverse morphologies [1, 2]. For instance, active contractile forces arising from cellular actin-myosin network can drive the in-plane or out-of-plane deformations of epithelial tissues, as observed in *Drosophila* mesoderm invagination [3], intestinal crypt formation [4], vertebrate neurulation [5, 6], and so on. Mechanical compression from neighboring tissue compartments is responsible for the undulated tissue surfaces of mucosa [7, 8], intestine [9-11], and brain cortex [12-14]. Hydrostatic pressure, arising from cellular osmotic regulation, has also been proven to power lumen formation and tissue expansion in a variety of settings [15-17].

Epithelial tissues have been found to show non-trivial deformation features. An interesting example is the body elongation of a marine animal *Trichoplax adhaerens*, whose self-motility drives remarkable morphological changes of its own epithelial tissue: the originally disk-like epithelial sheet forms holes and further breaks into thin, long threads [18]. Notably, with persistent stretching, the tissue thread becomes thinner in the middle and forms a necked region there, which propagates along the thread until the final breakage (Fig 1a). Similar deformation processes also occur *in vitro* [19-21]. Such tissue deformation behavior is reminiscent of the necking instability in metals [22], metallic glasses [23], glassy polymers [24] and soft elastomers [25]. Considère [26] first identified that the necking bifurcation of ductile materials occurred when the external load reached its maximum. Since then, the necking bifurcation of various materials has been studied extensively [27-31]. However, the necking phenomenon in soft tissues remains largely unexplored. Recent live-cell imaging shows that the finite deformation of epithelial tissues stems from individual events at the cellular scale [32], such as cell shape changes and cell rearrangements (Fig 1b). An epithelial monolayer is made of polygon-like cells that tightly bind via cell-cell junctions [5, 33]. Importantly, the cell edges can remodel (i.e. new edges can be created while existing ones can be annihilated) and the final cell shape is a result of the competition among the adhesion forces at the cell-cell junctions, fluid pressure of its cytoplasm, and contractile forces of its actomyosin networks near the cell-cell interface [5, 34]. Thus, to uncover the physical principle underlying the non-trivial tissue deformation like tissue necking, the individual cellular events should be taken into account.

Cell-based discrete tissue models, which capture basic mechanical and geometric properties of individual cells in soft tissues, have been developed to better understand how the mechanics and morphology of single cells affect tissue-scale deformation and morphogenesis. Representative cell-based models include the vertex model [35-39], cellular Potts model [40, 41], and particle-based models [42-44]. Among them, the vertex model is particularly attractive for the simulation of epithelial tissues, due to its ability to well capture the confluency of epithelial tissues, as well as key mechanical cues (such as cell adhesion and contractility) of individual cells and the remodeling of the cell-cell interface. The vertex modeling has successfully uncovered interesting features of epithelial tissues, such as cell packing irregularity [45] and fluid-solid rigidity transition [46]. However, most of the previous studies are limited to infinitesimal tissue deformation [47-50]. For instance, Bi [46] found that enhanced cell-cell adhesion (or weakened cell contractility) can drive a transition in which the tissue's instantaneous modulus softens to zero. On the other hand, there have been efforts to formulate the continuum models of epithelial tissues by combining ingredients from the vertex model and continuum field theories [47, 51-54]. With coarse-grained treatments, these continuum models capture averaged cell-scale mechanical evolutions via internal state variables, but ignore the details and the stochastic nature of individual cell events, thus failing to fully depict the versatile mechanical behaviors of epithelial tissues.

Here, a multiscale constitutive model is combined with discrete vertex simulations to investigate the necking instability emerging in epithelial sheets. The constitutive model is built upon two basic principles on the relationship between the tissue-scale deformation and cell events. First, a deformed cell can recover to its original shape after the removal of external loadings [55], thus the changes in cell shape contribute to tissue elastic deformation. Second, topological transition at the cellular scale generates inelastic tissue deformation [56, 57]: during the cellular topological transition, neighboring cells move apart and lose their contact, while previously unconnected cells intercalate and form a new cell-cell contact (Fig 1b). This paper is organized as follows. Section 2 introduces the cell-based tissue model (i.e. the vertex model) and the simulation protocol. In Section 3, we use the vertex model to simulate the uniaxial tensile test of an epithelial tissue, and reveal its necking instability at both the tissue scale and the cellular scale. In Section 4, the criteria for necking bifurcation and necking propagation are predicted based on the multiscale constitutive relation we propose, and validated by discrete vertex simulations. The effects of topological defects are discussed in Section 5. We summarize the main results drawn from this study in Section 6.

## 2. Cell-based tissue model

The mechanical behaviors of epithelial tissues can be understood from the mechanical interactions at the cellular scale, such as cell-cell adhesion and actomyosin-mediated tension along the cell edges. The vertex model is a type of multiscale mechanical model that links cell mechanics with tissue deformation. It treats each cell as an individual polygon containing several vertices, and each vertex represents a tri-cellular junction where cell edges meet, and on which force balance is written [37, 38, 45, 58-62].

The potential energy of the epithelial monolayer mainly arises from the mechanical resistance of cells to their shape changes, and would build up when the actual shapes of cells deviate from their preferred shapes. Given this, the potential energy of each cell can be written as a function of cellular morphometric parameters such as cell area and cell perimeter [46, 63]:

$$E_i = \frac{1}{2}[K_A(A_i - A_0)^2 + K_P(P_i - P_0)^2], \tag{1}$$

with $A_0$ (and $P_0$) the preferred cell area (and perimeter), $A_i$ (and $P_i$) the actual area (and perimeter) of cell $i$, and $K_A$ (and $K_P$) the rigidities of cell area (and perimeter). The first energy term in Eq (1) represents the area elasticity that originates from cell cytoskeleton, while the second energy term is contributed by the line tension of the cell edges (composed of cell membrane and cortical actomyosin filaments). For epithelial tissues composed of a single type of cell, it is reasonable to assume that all the cells have the same $A_0$ and $P_0$. Then the total energy of the epithelial tissue would be $E = \sum_i E_i$.

To simplify the numerical simulation and theoretical analysis, herein, we non-dimensionalize the cell energy by $K_A A_0^2$, cell area by $A_0$, and parameters of length (e.g. cell perimeter $P$ and spatial coordinates) by $\sqrt{A_0}$. Then the potential energy of cell $i$ can be non-dimensionalized as

$$e_i = \frac{E_i}{K_A A_0^2} = \frac{1}{2}[(a_i - 1)^2 + \kappa(p_i - \chi)^2], \qquad (2)$$

with $a_i = A_i/A_0$ and $p_i = P_i/\sqrt{A_0}$ respectively the dimensionless area and perimeter of cell $i$. The dimensionless energy (2) is regulated by the rigidity ratio $\kappa = K_P/(K_A A_0)$ and a dimensionless geometric parameter $\chi = P_0/\sqrt{A_0}$. The latter serves as the "shape index" of single cells [46], and has proven to be a crucial parameter that determines the initial tissue stiffness [45, 46].

In the vertex model, the changes in cell shape and arrangement are achieved through the movements of polygon vertices, which are driven by the potential forces applied to them. The potential force applied to vertex $\alpha$ is $\boldsymbol{F}_\alpha = -\sum_i \partial e_i/\partial \boldsymbol{r}_\alpha$, with $\boldsymbol{r}_\alpha$ the spatial position of vertex $\alpha$. This force should be balanced with the viscous force there, that is

$$\gamma \frac{d\boldsymbol{r}_\alpha}{dt} = -\sum_i \frac{\partial e_i}{\partial \boldsymbol{r}_\alpha}, \qquad (3)$$

with $\gamma$ the viscous coefficient, whose value would not affect the quasi-static problems studied in this work.

Treating each cell as an infinitesimal element in a continuum (i.e. the tissue), one can also introduce stress tensors to describe the mechanical state of single cells, which are subjected to discrete point forces acting on their vertices. For instance, the relationship between Cauchy stress $\boldsymbol{\sigma}$ and vertex forces can be established from an identical equation $\boldsymbol{\sigma} = (\boldsymbol{r} \otimes \boldsymbol{\sigma}) \cdot \nabla$. This equation holds for any second-order tensor that is symmetric and divergence-free. Integrating this equation over the cell area $a$ and applying the divergence theorem, we have

$$\int_a \boldsymbol{\sigma} \, da = \int_a (\boldsymbol{r} \otimes \boldsymbol{\sigma}) \cdot \nabla \, da = \oint_l \boldsymbol{r} \otimes \boldsymbol{\sigma} \cdot \boldsymbol{n} \, dl, \tag{4}$$

where $l$ stands for the cell boundary, and $\boldsymbol{n}$ is the unit vector normal to the cell boundary. Under the assumption of uniform stress within a cell, the mechanical state of cell $i$ can be represented by the Cauchy stress $\boldsymbol{\sigma}_i = a_i^{-1} \oint_l \boldsymbol{r}_i \otimes \boldsymbol{t}_i \, dl$, with $\boldsymbol{t}_i = \boldsymbol{\sigma}_i \cdot \boldsymbol{n}_i$ the force distributed at the cell boundary. In the vertex model, the boundary forces are discrete point forces that act on vertices associated with cell $i$, thus the Cauchy stress $\boldsymbol{\sigma}_i$ can be estimated as

$$\boldsymbol{\sigma}_i = \frac{1}{a_i} \sum_{\beta \in \text{cell } i} \boldsymbol{r}_\beta \otimes \boldsymbol{f}_{\beta i}, \tag{5}$$

with $\boldsymbol{f}_{\beta i} = \partial e_i / \partial \boldsymbol{r}_\beta$ the point force acting on the $\beta$-th vertex inside cell $i$. Submitting potential energy (2) into Eq (5), we obtain the specific expression of Cauchy stress $\boldsymbol{\sigma}_i$ as

$$\boldsymbol{\sigma}_i = (a_i - 1)\boldsymbol{I} + \frac{\kappa p_i (p_i - \chi)}{a_i} \boldsymbol{M}_i, \tag{6}$$

where $\boldsymbol{M}_i$ is a symmetric cell shape tensor defined as

$$\boldsymbol{M}_i = \frac{1}{p_i} \sum_{\beta \in \text{cell } i} l_{\beta,\beta+1} \boldsymbol{p}_{\beta,\beta+1} \otimes \boldsymbol{p}_{\beta,\beta+1}. \tag{7}$$

Here, $\boldsymbol{p}_{\beta,\beta+1}$ is the unit vector pointing from vertex $\beta$ to vertex $\beta + 1$, and $l_{\beta,\beta+1}$ is the dimensionless distance between these two vertices. See S1 Text for the derivation of Eqs (6) and (7).

## 3. Tissue necking instability

The epithelial tissue is modelled as a tiling of uniform hexagonal cells with unit area (Fig 2a). In our simulations, the total energy of the tissue is first minimized to obtain the stress-free state, from which we perform displacement-controlled uniaxial tensile tests. A displacement of 0.1% of the initial tissue length is imposed at each step, and the force balance Eq (2) for each vertex is solved using the forward Euler method [60, 64]. Unless otherwise stated, we set the initial tissue length $L = 30$, the initial tissue width $W = 10$, the rigidity ratio $\kappa = 0.16$, and the shape index $\chi = 3.7$.

Importantly, and consistent with experimental observations (Fig 1a), the uniaxial stretching of the epithelial tissue naturally leads to necking (Fig 2a): with continued stretching, the tissue first undergoes homogeneous deformation, then locally narrows, and this narrow region gradually propagates and causes the tissue body to separate into two regions with distinct widths. In the region with a marked decrease in tissue width (i.e. the necked region), all the cells have their neighbors changed, and the cell edges along the loading direction undergo zigzag-to-armchair transition. On the other hand, cells in the less-deformed region change their shapes but not their neighbors, indicating pure elastic deformation in this region. The necked region can steadily propagate along the tissue, while the tissue width (and cell shape) in both the necked and un-necked regions remains constant (Fig 2a). During the necking propagation, the nominal tissue stress $s$ drops to a nearly constant level (Fig 2b). Once the necked region propagates through the whole tissue, the tissue becomes homogeneous again and the tissue stress $s$ rises up until the final catastrophic failure. The stress-stretch curve (Fig 2b) clearly shows the four stages of tissue deformation: i) initial homogeneous elastic deformation, ii) steady necking propagation after bifurcation, iii) tissue stiffening, and iv) final catastrophic failure.

The necking front of the epithelial tissue undergoes inelastic shear deformation (Fig 3), highly reminiscent of the plastic flow in metallic materials [65]. Interestingly, the tissue necking front also forms "slip lines" similar to those observed in metals. The slip lines are formed by continuous topological transition events (Fig 3): as the necked region propagates forward, the cells in front of the necked region are forced to move inward and change their positions and neighbors, and such transition events occur successively along an oblique line. Cellular topological transition rearranges the cell positions and narrows the tissue, and thus promotes the necking propagation in the epithelial tissue. Besides topological transition, the shapes of these cells are also distorted and the shear stress $\tau$ accumulates in the necking front (Fig 3).

## 4. Theoretical analysis

As a limiting-point instability, the necking of materials is initiated when the load reaches the peak value in the uniaxial stress-stretch curve [26].On the other hand, the steady necking propagation is analogous to "phase-separation": the necked region and the un-necked region coexist and evolve with uniaxial stretching [22, 66]. The constitutive relationship, i.e. the uniaxial stress-stretch relation of materials undergoing homogeneous deformation, has been shown to provide key information on both the necking bifurcation and the necking propagation. Therefore, in the following section, we first derive the uniaxial stress-stretch relation of epithelial tissues with homogenous deformation, then use this relation to predict the bifurcation condition for tissue necking and the tissue mechanical state during necking propagation. Comparisons between the theoretical predictions and numerical simulations are also given.

### 4.1. Multiscale constitutive model

Consistent with the discrete vertex simulations in Section 3, we model the tissue as a tiling of initially uniform and regular hexagonal cells. We take the initial stress-free configuration as the reference configuration, then analyze the tissue elasticity that involves changes in cell shape, and also the inelastic tissue deformation arising from topological transition at the cellular scale.

### 4.1.1. Initial stress-free configuration

In the stress-free state, cells are fully relaxed and have minimal potential energy. In each hexagonal cell, all the edges have the same length $d_I$ and all the interior angles are equal to $\theta_I = 120°$ (Fig 4a), thus the cellular potential energy (2) can be simplified as $e_I(d_I) = (\frac{3\sqrt{3}}{2} d_I^2 - 1)^2 + \kappa(6d_I - \chi)^2$. Note that the energy $e_I$ becomes zero when the shape index $\chi$ reaches the critical value $\chi^* = 2\sqrt{6\tan\frac{\pi}{6}} \approx 3.72$. In this scenario, we have the cell perimeter $p_I = \chi^*$ and the cell area $a_I = 1$. Otherwise, we can introduce $k = 6d_I/\chi^*$, which is the relative edge length with respect to the ground state. This relative length $k$ should satisfy $de_I/dd_I = 0$, which suggests

$$2k(k^2 - 1) + \chi^*\kappa(k\chi^* - \chi) = 0. \tag{8}$$

And we have the cell area $a_I$ and the cell perimeter $p_I$ as

$$a_I = k^2, \quad p_I = k\chi^*. \tag{9}$$

We can also characterize the initial cell shape in terms of the cell length $l_I$ and cell width $w_I$ (Fig 4a). They are related to the edge length $d_I$ and angle $\theta_I$ as $l_I = 2d_I \sin\theta_I$, $w_I = d_I(1 - \cos\theta_I)$, and can be expressed as

$$l_I = \frac{k\chi^*}{2\sqrt{3}}, \quad w_I = \frac{k\chi^*}{4}. \tag{10}$$

### 4.1.2. Tissue elasticity: cell deformation

Subjected to external forces, the epithelial tissue will deform. Refer to the initial stress-free configuration, we can define $\lambda$ and $\lambda'$ as the tissue stretch ratios respectively along the length and width directions. In the absence of cell rearrangement, cell shape changes are synchronized with the tissue deformation, which means the current cell length and width respectively become $\lambda l_I$ and $\lambda' w_I$ (Fig 4a), and the current cell area $a$ and perimeter $p$ yield

$$a = \lambda \lambda' a_I = \lambda \lambda' k^2,$$
$$p = 2\lambda' w_I + f(\theta)\lambda l_I = \left[\frac{f(\theta)\lambda}{\sqrt{3}} + \lambda'\right]\frac{k\chi^*}{2},$$
(11)

with $f(\theta) = (2 + \cos\theta)/\sin\theta$. Submitting Eq (11) into Eq (2), we can obtain the potential energy of a single cell as

$$e(\lambda, \lambda', \theta) = \frac{1}{2}(\lambda\lambda' k^2 - 1)^2 + \frac{\kappa}{2}\left\{\left[\frac{f(\theta)\lambda}{\sqrt{3}} + \lambda'\right]\frac{k\chi^*}{2} - \chi\right\}^2.$$
(12)

Besides two stretch ratios $\lambda$ and $\lambda'$, the energy (13) is also dependent on the interior angle $\theta$. In a regular cell lattice, the tissue energy density equals the energy of a single cell divided by its initial area in Eq (9), that is $e/k^2$. Then, the nominal tissue stresses can be written as

$$s = \frac{\partial e(\lambda, \lambda', \theta)}{k^2 \partial \lambda}, \quad s' = \frac{\partial e(\lambda, \lambda', \theta)}{k^2 \partial \lambda'}.$$
(13)

As the angle $\theta$ can adjust freely, the energy minimization requires

$$\frac{\partial e(\lambda, \lambda', \theta)}{\partial \theta} = 0.$$
(14)

Eq (14) suggests $f'(\theta) = 0$, that is $\theta = 120°$. This means epithelial cells always undergo isogonal deformation [67]. Therefore, the nominal stresses (13) yield

$$s = \lambda'(k^2\lambda\lambda' - 1) + \frac{\kappa\chi^*[-2\chi + k\chi^*(\lambda + \lambda')]}{4k}, \quad (15)$$

$$s' = \lambda(k^2\lambda\lambda' - 1) + \frac{\kappa\chi^*[-2\chi + k\chi^*(\lambda + \lambda')]}{4k}. \quad (16)$$

For the uniaxial tensile test (Fig 2), the transverse stress $s' = 0$ (14) and the longitudinal nominal stress $s$ can be determined as (Fig 4b)

$$s = \frac{2\kappa\chi^*(\chi^* - 2k\chi\lambda + k^2\chi^*\lambda^2)[\kappa\chi^*(-\chi + k\chi^*\lambda) + 2k\lambda(-1 + k^2\lambda^2)]}{k(\kappa\chi^{*2} + 4k^2\lambda^2)^2}. \quad (17)$$

In discrete vertex simulations (Fig 2b), we have the nominal stress $s = \sigma\lambda'$, with $\sigma$ the uniaxial Cauchy stress calculated from Eq (6). The theoretical prediction of the tissue stress (17) well fits the simulation results in various parameter settings (Fig 4c and 4d). We also find that the shape index $\chi$ and the rigidity ratio $\kappa$ show opposite effects on tissue stiffness: when the shape index $\chi$ increases and gets closer to the critical value $\chi^*$, the single cells have a smaller potential energy (see Subsection 4.1.1 for details) and thus become easier to deform; while an increase in the rigidity ratio $\kappa$ would increase the line tension at the cell edges, raise the cellular potential energy (12), and thereby stiffen the cells.

### 4.1.3. Tissue inelasticity: cell topological transition

Next, we examine the tissue inelastic deformation arising from topological transition (i.e. cell rearrangement) (Fig 5a). The topological transition occurs through the annihilation and creation of cell edges: the cell edges perpendicular to the loading direction become shorter during tissue deformation, and when its length $d = (3\lambda' - \lambda)kd_I/2$ drops below the critical threshold $d_T$ (set as 0.1 in this study), a new cell edge parallel to the loading direction will form and elongate. Such topological transition rearranges epithelial cells.

Denote $\lambda_T$ as the critical stretch ratio when the topological transition occurs. It should satisfy the critical condition $d = d_T$, that is

$$\chi^*[6k\lambda_T - 2k^3\lambda_T^3 + \kappa\chi^*(3\chi - 2k\chi^*\lambda_T)] = 6(\kappa\chi^{*2} + 4k^2\lambda_T^2)d_T. \tag{18}$$

The topological transition events affect the cell shape and cause discontinuous changes in cellular deformation. At the topological transition, the tissue stretch ratio $\lambda_T$ remains unchanged, but the cell shape may change due to cell rearrangements, and the corresponding cellular stretch ratio becomes different from the tissue stretch ratio $\lambda_T$. Let $\lambda_T^-$ and $\lambda_T^+$ respectively be the cellular stretch ratios right before and after the transition point. Given that the tissue deformation is purely elastic before transition, we immediately have $\lambda_T^- = \lambda_T$. After topological transition, however, cells change positions and the cellular stretch ratio $\lambda_T^+$ is no longer the same as the tissue stretch ratio $\lambda_T$, i.e. $\lambda_T^+ \neq \lambda_T$. Given that the total length of a four-cell unit remains constant at transition (see schematics B and C in Fig 5a) and this cell unit now has three cells instead of two along the length direction, we can establish the relationship between two cellular stretch ratios as $2\lambda_T^- l_I = 3\lambda_T^+ w_I$, which suggests

$$\lambda_T^+ = n\lambda_T, \text{with } n = \frac{4\sqrt{3}}{9}. \tag{19}$$

One can easily find $\lambda_T^+ < \lambda_T^- = \lambda_T$, indicating that the topological transition mitigates cellular shape changes. Specifically, the cell rearrangement inserts new cells into the load-bearing axis, which accommodates the tissue-level elongation among a greater number of cells. This effectively reduces the deformation and, consequently, the stress within each cell.

Further tissue deformation after the topological transition still arises from changes in cell shape. As the ratio of current tissue length to the tissue length at the transition point is

$\lambda/\lambda_T$, we obtain the stretch ratio of single cells as $\lambda_T^+(\lambda/\lambda_T) = n\lambda$. Similar to the analysis in Subsection 4.1.2, we obtain the tissue energy density $e(\lambda, \lambda', \theta)$ and corresponding nominal stress $s(\lambda)$ with $\lambda > \lambda_T$ (see S2 Text for details). The complete stress-stretch relation including topological transition can be summarized as

$$s = \begin{cases} \dfrac{2\kappa\chi^*(\chi^* - 2k\chi\lambda + k^2\chi^*\lambda^2)[\kappa\chi^*(-\chi + k\chi^*\lambda) + 2k\lambda(-1 + k^2\lambda^2)]}{k(\kappa\chi^{*2} + 4k^2\lambda^2)^2} & \lambda < \lambda_T \\ \dfrac{2\kappa\chi^*[\chi^* - 2k\chi n\lambda + k^2\chi^*(n\lambda)^2]\{\kappa\chi^*(-\chi + k\chi^* n\lambda) + 2kn\lambda[-1 + k^2(n\lambda)^2]\}}{\sqrt{3}k[\kappa\chi^* + 4k^2(n\lambda)^2]^2} & \lambda > \lambda_T \end{cases}. \quad (20)$$

This stress-stretch relation is discontinuous as the topological transition causes stress relaxation of the tissue (Fig 5b).

## 4.2. Analysis of tissue necking

The necking bifurcation occurs when the tissue stress reaches its peak value [22, 26]. From the discontinuous constitutive relation (20), the necking bifurcation is coincident with the topological transition. This means the tissue stretch ratio at bifurcation is just $\lambda_T$ in Eq (19) and the bifurcation stress is $s_T = s(\lambda_T)$ in Eq (18).

After the onset of necking, the tissue will enter the stage where the necked phase steadily propagates forwards (Fig 2a). In this stage, the nominal stress remains constant at a value $s_*$ (Fig 2b). To analyze this process, we consider an infinitesimal tissue section (with initial length $\mathrm{d}L$) in front of the neck. When the neck front shifts forward to engulf this section, this section undergoes significant deformation with a length increase of $(\lambda_N - \lambda_U)\mathrm{d}L$, where $\lambda_N$ and $\lambda_U$ are respectively the stretch ratios in the necked and un-necked states (Fig 6a). The work done is therefore $s_*(\lambda_N - \lambda_U)\mathrm{d}L$. This work should be equal to $\mathrm{d}W = \mathrm{d}L \int_{\lambda_U}^{\lambda_N} s(\lambda)\mathrm{d}\lambda$, which is the energy increase of this section as it passes from the un-necked state to the necked state. Then the propagation stress $s_*$ should yield

$$s_*(\lambda_N - \lambda_U) = \int_{\lambda_U}^{\lambda_N} s(\lambda)d\lambda = \int_{\lambda_U}^{\lambda_T} s(\lambda)d\lambda + \int_{\lambda_T}^{\lambda_N} s(\lambda)d\lambda, \qquad (21)$$

where $s(\lambda)$ is given in Eq (20). Eq (21) has a simple graphical explanation (Fig 6b): the rectangular area $s_*(\lambda_N - \lambda_U)$ is equal to the area under the stress-stretch curve $s(\lambda)$ in the interval from $\lambda_U$ to $\lambda_N$, thus the areas of the two lobes (i.e. the shaded regions in Fig 6b) are also equal. This equality is also known as Maxwell's condition for the coexistence of two phases [66, 68]. Once the propagation stress $s_*$ is determined, one can also obtain the stretch ratios $\lambda_U$ and $\lambda_N$ from the stress-stretch relation (20) (Fig 6b).

Our theoretical predictions well fit the simulation results in various parameter settings, and reveal the dependence of tissue necking behaviors on tissue properties (such as the shape index $\chi$) (Fig 7a). As mentioned in Subsection 4.1.2, an epithelial tissue with a larger shape index $\chi$ is mechanically softer, thus the bifurcation stress $s_T$ and propagation stress $s_*$ also become smaller with the increase in $\chi$ (Fig 7b and 7c). On the other hand, the tissue stretch ratios, represented by $\lambda_N$ (for necked region) and $\lambda_U$ (for un-necked region), are insensitive to the shape index $\chi$ (Fig 7d and 7e). In our theoretical analysis based on the uniaxial stress-stretch relation, the bifurcation condition is independent of the initial tissue length or width, which is also consistent with the vertex simulation results (Fig 8a and 8b): the bifurcation stress $s_T$ remains close to the theoretical prediction when the tissue length or width varies. We notice that, as the initial tissue width $W$ increases, the bifurcation transforms from a symmetric mode to an anti-symmetric mode (Fig 8a), the latter leads to localized shearing instead of necking [69]. Fig 8c shows that the symmetric bifurcation and corresponding necking instability always occur unless the initial tissue width $W$ becomes larger than the tissue length $L$, and the values of the propagation stress $s_*$ obtained from vertex simulations consistently align with the theoretical prediction (21). Overall, the validation of the theoretical predictions demonstrates that our theoretical model efficiently captures the general features of the necking of epithelial tissues.

## 5. Effects of topological defects

In previous sections, epithelial cells are assumed to be well-organized into regular hexagonal shapes. This assumption enables to derive simple constitutive descriptions of the epithelial tissue, and yields key predictions regarding tissue necking. In real epithelial tissues, however, cell shapes can be irregular and the cell arrangement is also disordered [45, 57, 70], see also Fig 1b for the epithelium of a *Drosophila* embryo [32]. A notable feature is the presence of epithelial cells with fewer or more than six edges, resembling pentagons or heptagons rather than hexagons. Typically, a pentagonal cell pairs with a heptagonal one, forming a pentagon-heptagon defect pair [71-73]. Analogous to two-dimensional carbon materials like graphene [74], these non-hexagonal cells can act as topological defects and alter tissue deformation [60, 75]. Given these, we further investigate the influence of topological defects on tissue necking by performing vertex simulations. We find that, in the elastic deformation stage, the nonlinear stress-stretch curves of tissues with topological defects closely match those of ordered tissues (i.e. hexagonal cell lattice) (Fig S1), indicating that tissue elasticity is insensitive to topological defects. The inelastic tissue deformation and the tissue necking behavior, however, can be affected by topological defects, as detailed below.

To evaluate the impact of topological defects on tissue necking, we first embed a pentagon-heptagon pair into a regular cell lattice to form a dispersed "point defect" (Fig

9a). Such topological pairs are widely observed in epithelial tissues across diverse species [72]. We find that even a single point defect is sufficient to trigger the tissue neck to initiate from the site where the defect is located (Fig 9b), indicating that the necking bifurcation of tissues is highly sensitive to topological defects. Such high sensitivity to material imperfections is also a characteristic of the necking of metallic materials [76]. In contrast, this defect pair has little impact on the overall stress-stretch curve (Fig 9c).

Next, we generate disordered cell lattices using Voronoi tessellation [37, 62, 77]. The randomness of topological defects leads to diverse necking behaviors across different tissue samples. The number of topological transition events (blue dotted line in Fig. 10) increases after the bifurcation and is synchronized with the fluctuations of the stress curve (red line in Fig. 10), where pronounced stress drops occur due to cell rearrangements. Following the necking bifurcation, some tissue samples undergo catastrophic failure directly, without necking propagation (e.g. Sample I in Fig 10a), whereas in others (e.g. Sample II in Fig 10b), the neck region propagates persistently, as in ordered tissues (i.e. regular cell lattice in Fig 2). Apparently, the second mode exhibits higher ductility and toughness than the first one. Both necking modes can be observed in tissue samples with diverse shapes and sizes (see Fig S2 for details, where the tissue aspect ratio $L/W$ varies from 1 to 4 and the total cell number $N = 150, 300, 600$), and in both cases, the necked region collapses into a thin thread before final rupture, reminiscent of the morphological evolution of *Trichoplax adhaerens* (Fig 1a). In the second necking mode, the disordered tissue may develop multiple necked regions that propagate towards each other, a phenomenon also observed in experiments (Fig 10b).

The bifurcation stress $s_T$ remains consistent across disordered tissue samples, albeit at a lower level than in ordered tissues (Fig 11a). This indicates that initial defects facilitate necking bifurcation. Although the occurrence and extent of necking propagation can be greatly affected by topological defects (Fig 9 and 10), the necking propagation stress $s_*$ across disordered samples remains close to that of ordered tissues (Fig 11b). The maximum stretch ratio, defined as the stretch at which the tissue undergoes complete failure, varies across samples; however, for most tissues it is clustered around a value of approximately 2.0. Finally, we find that necking bifurcation in disordered tissues is governed by tissue properties in a manner analogous to ordered tissues (Fig 11c and 11d). This suggests that our parameter analysis on ordered tissues (see Section 4 for details) provides a qualitative framework applicable to disordered systems.

## 6. Conclusion

In this article, we integrate discrete vertex simulations with theoretical modeling to unravel the necking instability of epithelial tissues. We propose a multiscale constitutive model that bridges the mechanical events of single cells and tissue-scale deformation. Specifically, the cellular shape evolution corresponds to elastic tissue deformation, while topological transitions of single cells govern inelastic tissue deformation. Based on this constitutive model, we predict the bifurcation condition for tissue necking and the steady state of necking propagation. Vertex simulations quantitatively validate our theoretical predictions. We also extend our tissue model from ordered cell lattices to disordered ones, and find that initial topological defects influence the necking bifurcation and impede the necking propagation. Our simulations of disordered tissues recapitulate the necking morphologies of real epithelial tissues. Beyond necking, our work also provides valuable insights for understanding other large-scale deformation behaviors of epithelial tissues, such as the gastrulation [78] and body axis elongation [32] of embryos, and the morphogenesis of skin cancers [79].

In the future, our multiscale constitutive model can be extended to complex loading scenarios [50]. Additional mechanical characteristics of epithelial cells can also be included in the model to more accurately recapitulate the large-scale tissue deformations observed *in vivo*. For instance, it has been revealed that cells can actively adjust their shapes, and even push and pull on one another to create internal forces that trigger large-scale deformation [52, 62, 80, 81]. Moreover, fundamental physiological processes such as cell division and apoptosis have been recognized to affect the tissue mechanical properties and drive tissue morphogenesis [82-85]. Addressing these issues will require the development of refined theories in tissue mechanics.

(a) Tissue necking

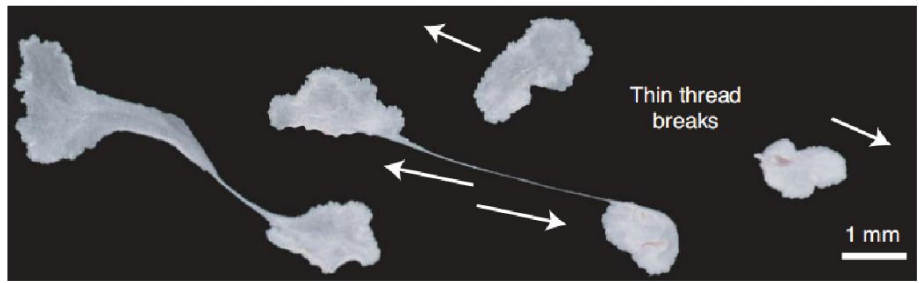

(b) Multiscale deformation

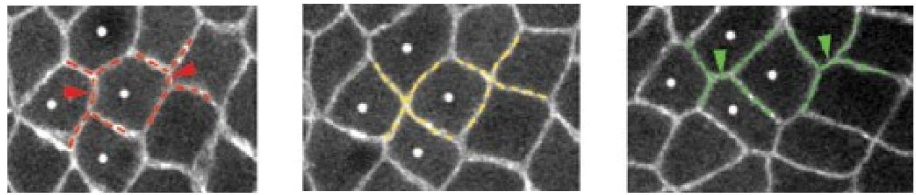

**Fig 1. Tissue necking and cellular topological transition. (a) Necking instability during the body elongation of *Trichoplax adhaerens*, a simple marine animal mainly composed of epithelial tissues (Prakash et al., 2021). From left to right, the necked region gradually propagates until final breakage. (b) Representative topological transition event at the cell scale in the epithelial tissue from a *Drosophila* embryo (Bertet et al., 2004): in the four-cell unit, the cell-cell junctions (red arrows) first shrink into points, then elongate perpendicular to the original direction (green arrows). After this transition, cells change their positions and neighbors.**

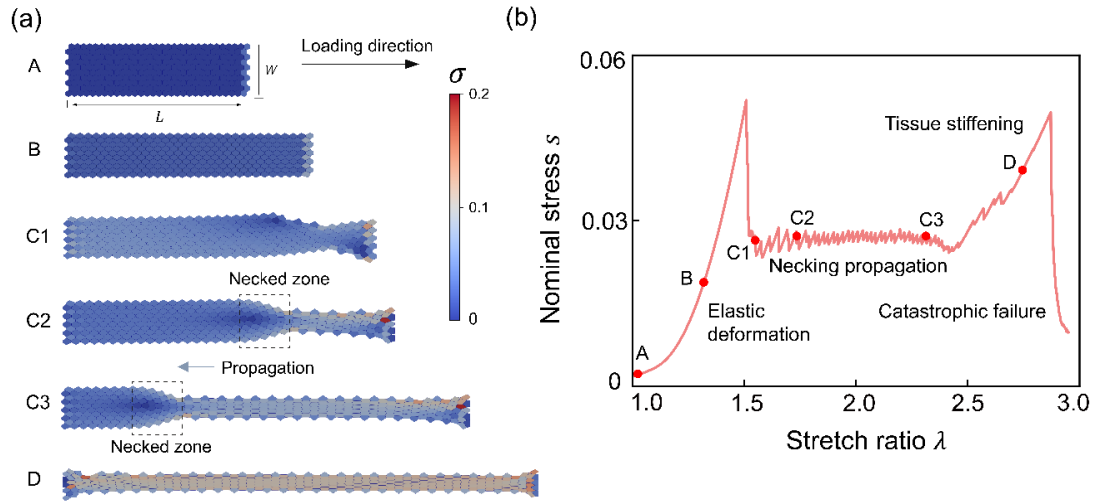

**Fig 2. Discrete vertex simulation of uniaxial tissue stretching. (a)** Representative snapshots of the tissue at different stages: **(A)** initial stress-free state, **(B)** elastic deformation, **(C1-C3)** necking bifurcation and propagation, and **(D)** tissue stiffening. The spatial distributions of the uniaxial Cauchy stress $\sigma$ are shown in each snapshot. **(b)** The resulting nonlinear stress-stretch response of the epithelial tissue under uniaxial tension.

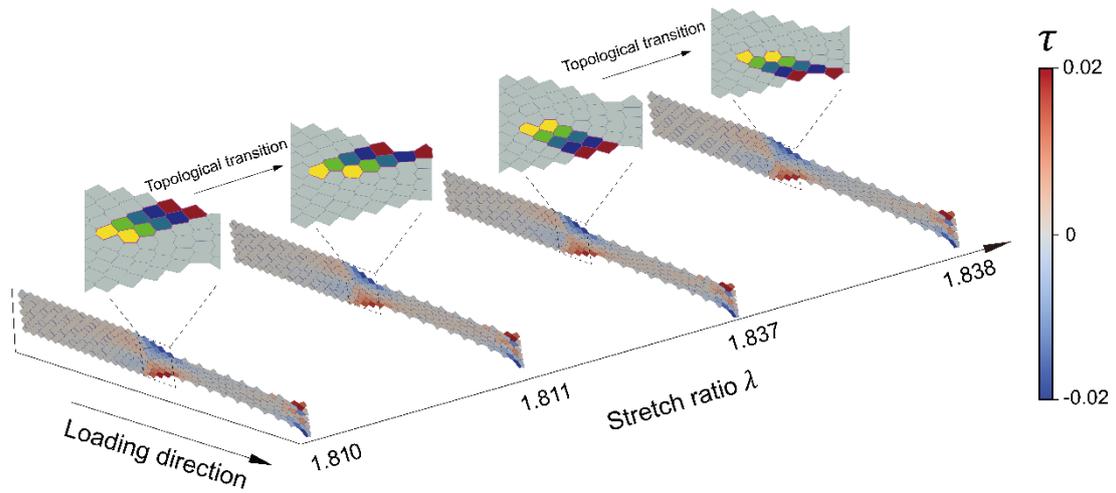

**Fig 3. Tissue necking is accompanied by topological transitions at the cell scale.** During the necking propagation, neighboring cells labeled with the same color undergo topological transitions and separate from each other. Both cellular topological transitions and shear stress $\tau$ are localized to the necking front.

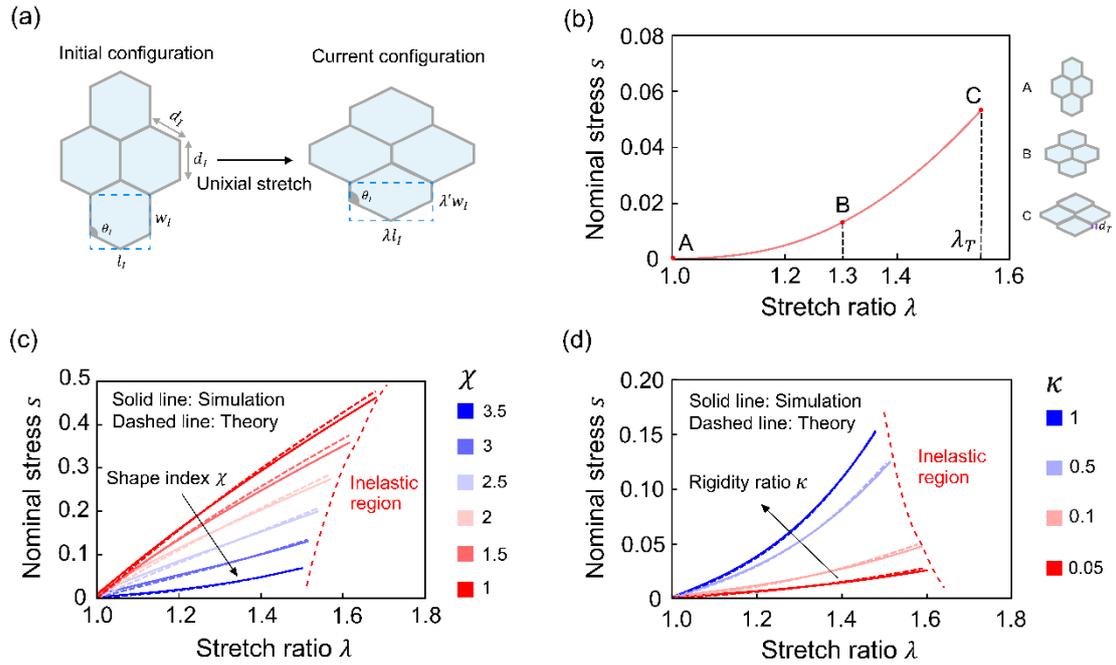

**Fig 4. Tissue elastic deformation with changes in cell shape.** (a) The initial and current configurations of a four-cell unit in a regular cell lattice. The left schematic defines the initial cell geometry, including the edge length $d_I$, interior angle $\theta_I$, cell length $l_I$ and cell width $w_I$. The right schematic shows the current cell length $\lambda l_I$ and width $\lambda' w_I$, with $\lambda$ and $\lambda'$ the tissue stretch ratios. (b) Theoretical prediction of the tissue stress-stretch curve before topological transition (i.e. $\lambda < \lambda_T$, with $\lambda_T$ the critical stretch ratio for topological transition), shown with representative unit configurations. Discrete vertex simulations and theoretical predictions of the tissue stress-stretch curves for various (c) shape index $\chi$ and (d) rigidity ratio $\kappa$.

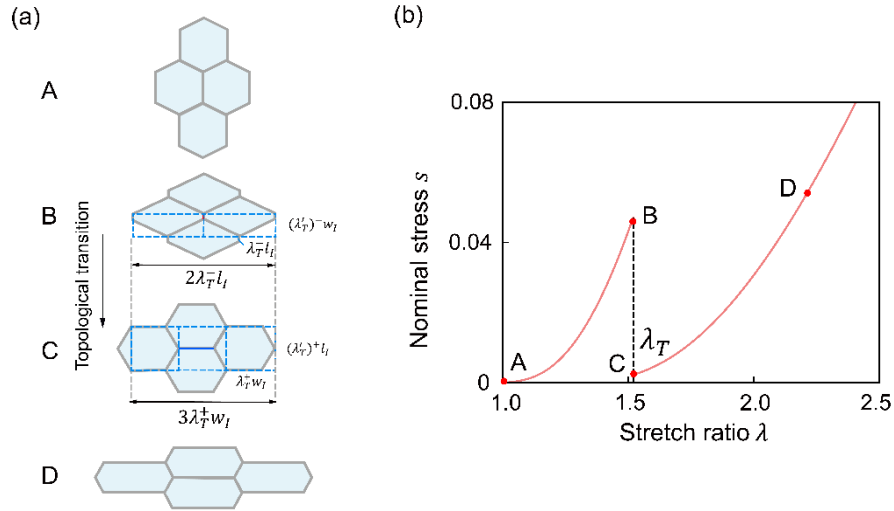

**Fig 5. Tissue inelastic deformation with cellular topological transition. (a) Schematics of a four-cell unit that undergoes topological transition. At the transition, the overall length of the cell unit remains constant, but the cell shape changes abruptly: the longitudinal cell stretch jumps from $\lambda_T^-$ to $\lambda_T^+$, while the transverse stretch jumps from $(\lambda_T')^-$ to $(\lambda_T')^+$. (b) Theoretical stress-stretch relation of an epithelial tissue with homogeneous deformations, where tissue inelasticity arises from cellular topological transition that occurs at $\lambda = \lambda_T$.**

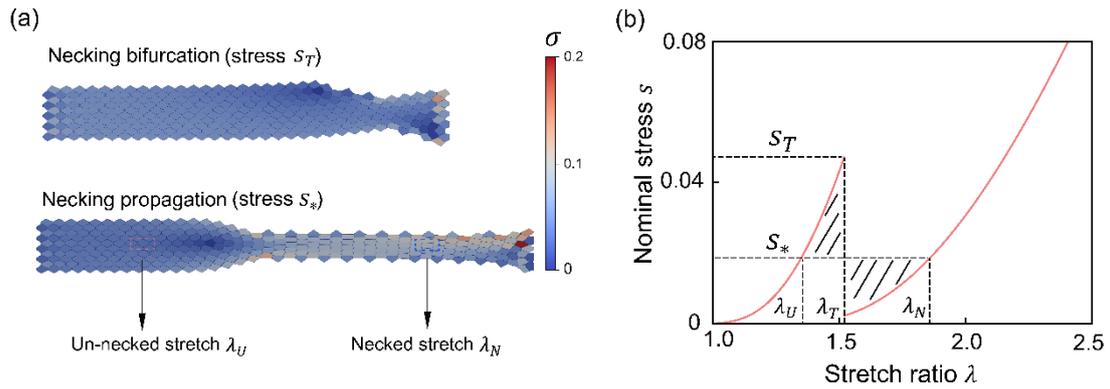

**Fig 6. Necking predictions based on the constitutive relation. (a) Simulation snapshots for (upper) necking bifurcation and (lower) necking propagation. The second snapshot shows the coexistence of the necked region with stretch $\lambda_N$ and the un-necked region with stretch $\lambda_U$. (b) The theoretical stress-stretch relation (for homogeneously deformed tissues) reveals the bifurcation stress $s_T$ and the mechanical state during necking propagation, characterized by the propagation stress $s_*$ and two stretch ratios $\lambda_U$ and $\lambda_N$. The shaded regions have equal areas.**

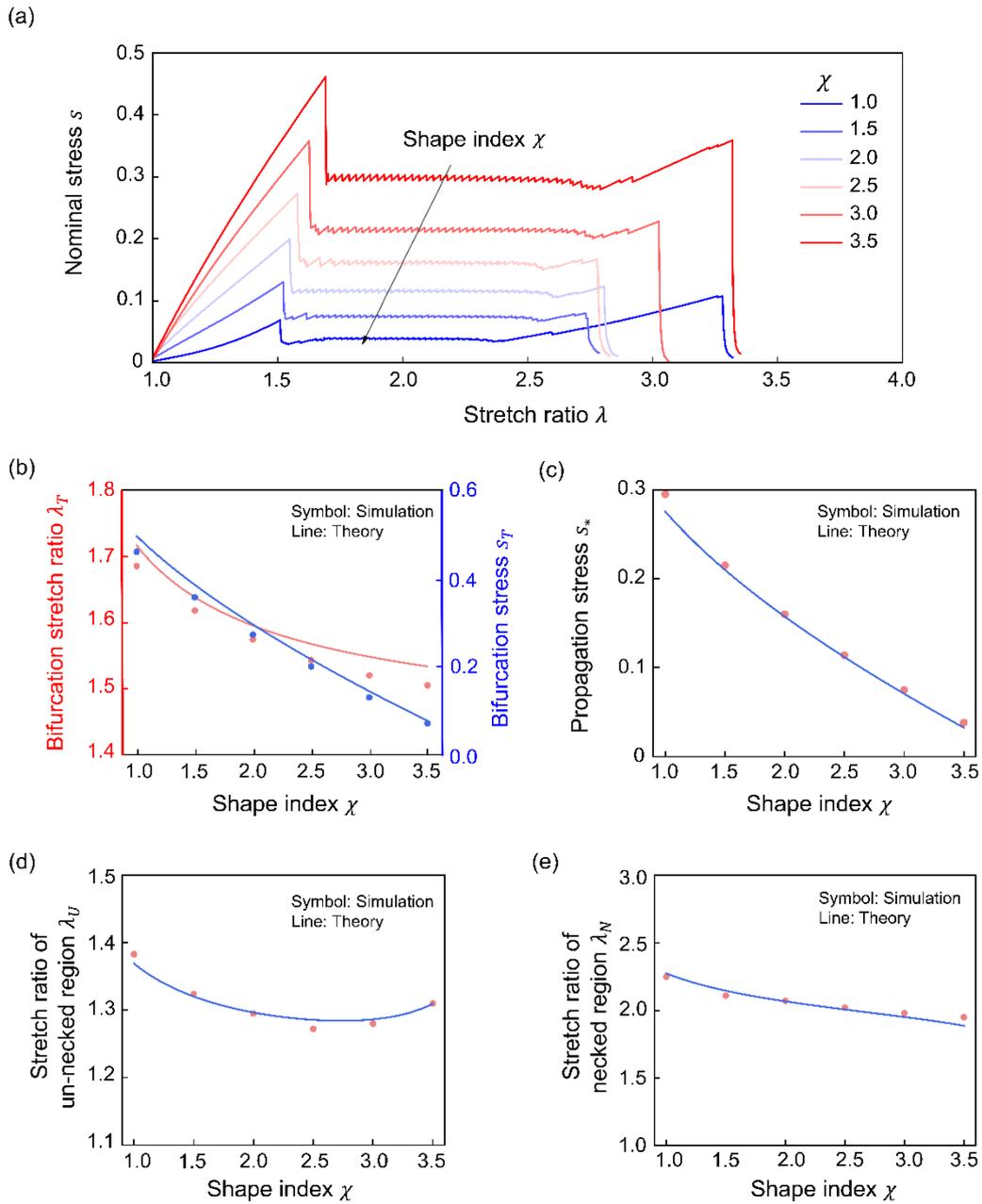

Fig 7. Theoretical predictions and vertex simulations of tissue necking with various shape indices $\chi$. (a) Simulation results of the uniaxial stress-stretch curves for different values of $\chi$. Comparison of theoretical predictions with simulation results for the (b) bifurcation stress $s_T$ and corresponding stretch ratio $\lambda_T$, (c) propagation stress $s_*$, (d) stretch ratio of the un-necked region $\lambda_U$, and (e) stretch ratio of the necked region $\lambda_N$.

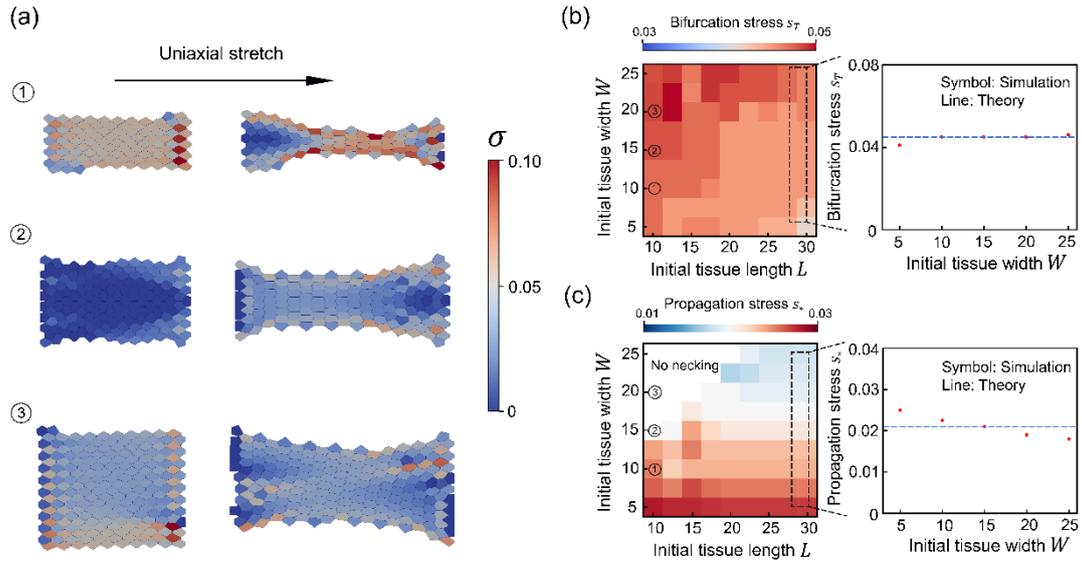

**Fig 8. Influence of tissue dimensions on tissue necking.** (a) Simulation snapshots of epithelial tissues with different initial widths $W$ (with the initial length $L = 10$). Stretch ratios in the first and second snapshots are respectively 1.4, 2.1. Phase diagrams illustrating the influence of tissue length $L$ and width $W$ on the (b) bifurcation stress $s_T$ and (c) propagation stress $s_*$. As the tissue width $W$ increases (with $L = 30$), both the bifurcation stress $s_T$ and propagation stress $s_*$ remain close to theoretical predictions.

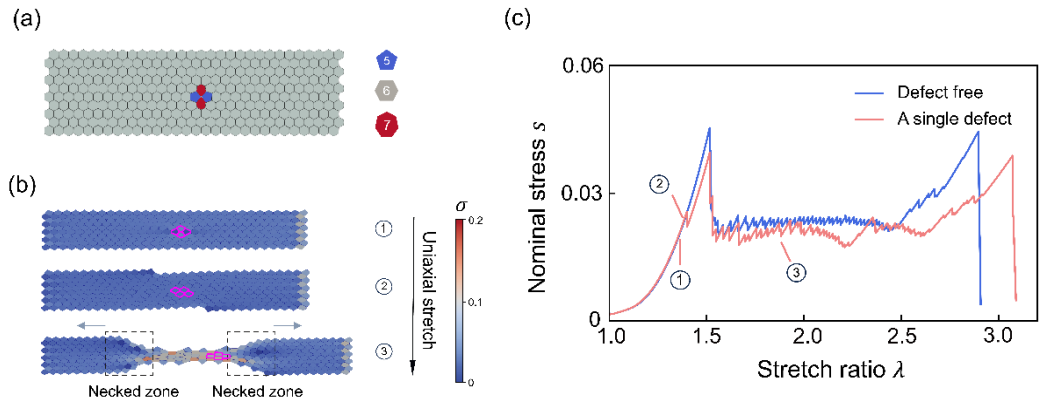

**Fig 9. Necking behavior of the epithelial tissue with a pentagon-heptagon defect pair. (a) The initial tissue configuration and (b) deformed configurations, showing that necking initiates from the defect site. (c) Corresponding uniaxial stress-stretch response, compared to that of a defect-free tissue (see also Fig 2b).**

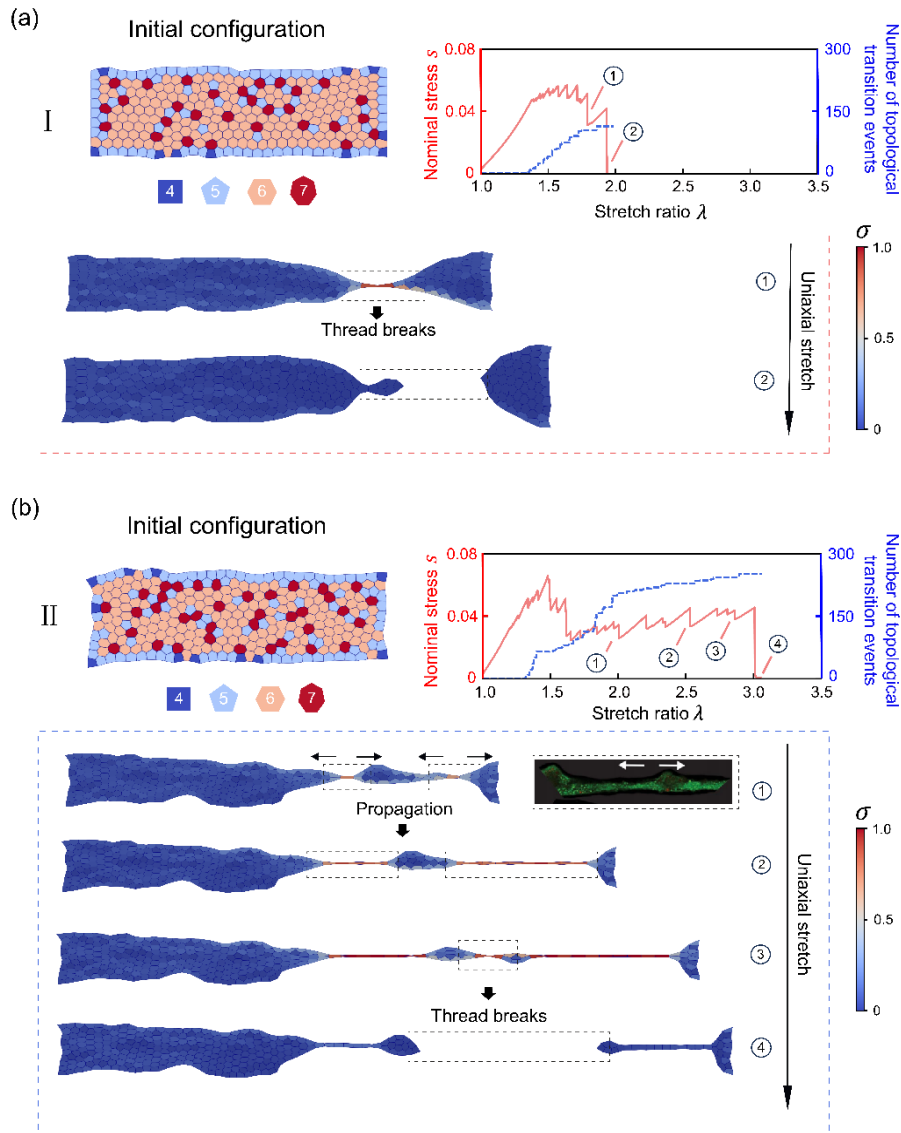

**Fig 10. Necking behaviors of disordered tissues. Disordered tissues exhibit two necking modes: (a) catastrophic failure without necking propagation; (b) failure preceded by persistent necking propagation. For each mode, the panels display a representative initial configuration alongside the evolution of the nominal stress $s$ and the cumulative number of topological transition events as a function of the tissue stretch ratio $\lambda$. The shape index is $\chi = 3.5$ in simulations.**

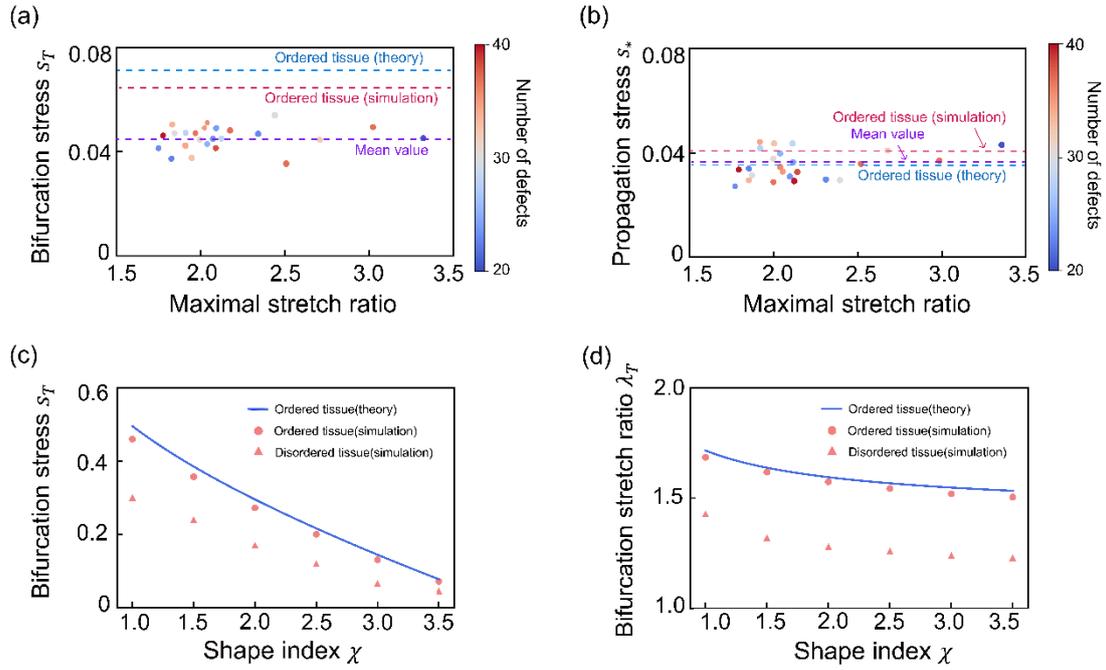

Fig 11. Necking stresses of disordered tissues. (a) The bifurcation stress $s_T$ and (b) propagation stress $s_*$ of disordered tissues with random topological defects, where the symbol color represents the number of initial defects and the mean stress values (dashed lines) are compared with those of the ordered (defect-free) tissues. Dependence of the (c) bifurcation stress $s_T$ and (d) bifurcation stretch $\lambda_T$ on the shape index $\chi$ for both disordered and ordered tissues.

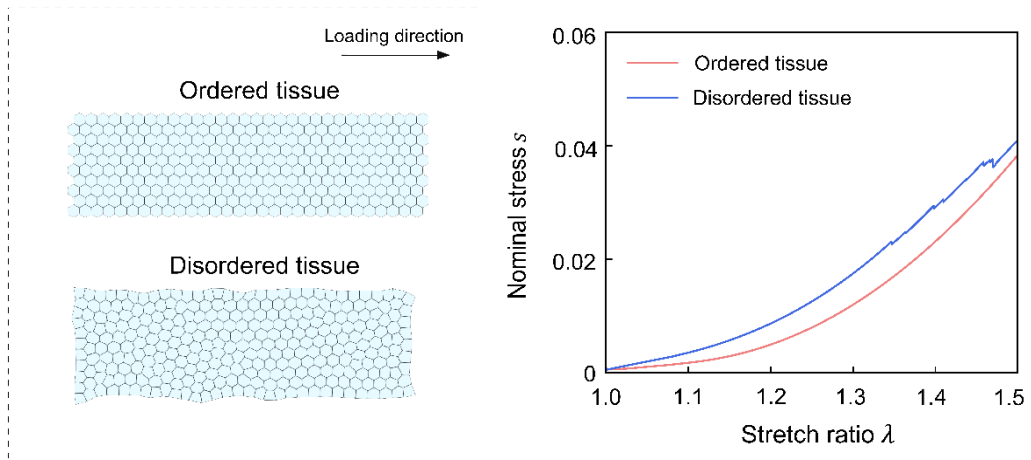

**Fig S1.** Comparison of elastic stress-stretch curves between ordered and disordered tissues.

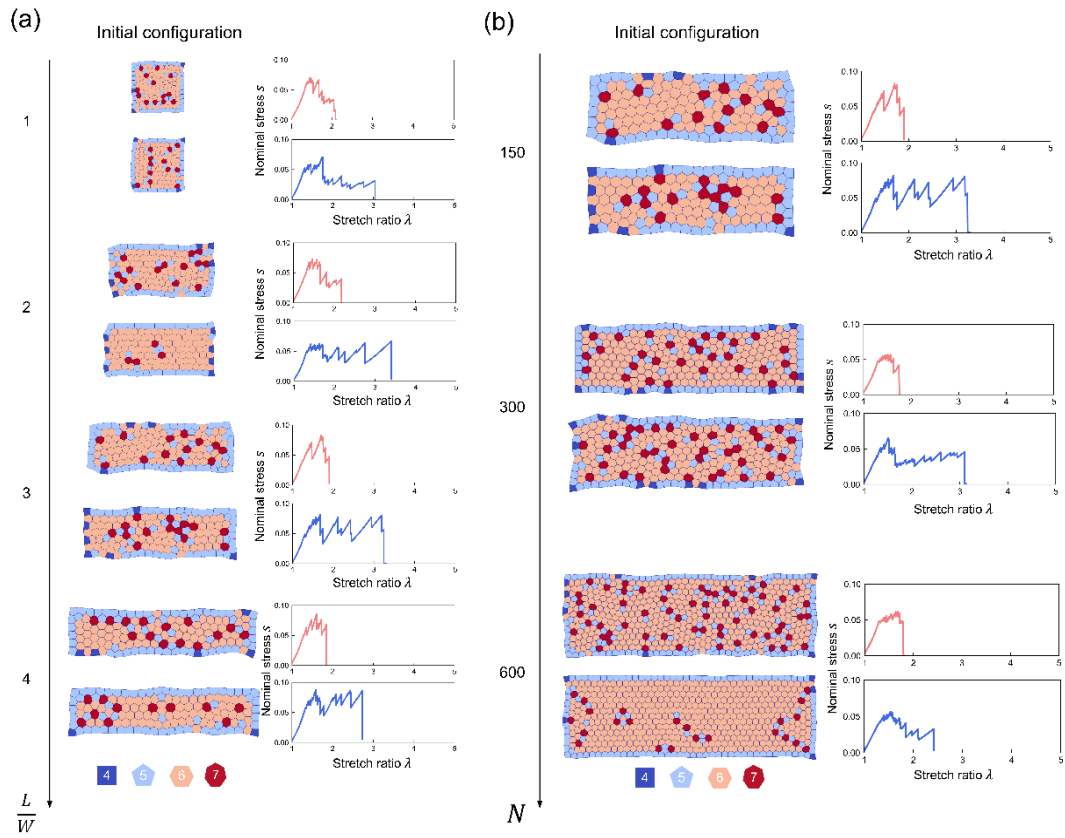

Fig S2. Two necking modes are observed in disordered tissues across a range of geometric parameters: (a) various aspect ratios $L/W$ with a fixed cell number $N = 150$, and (b) various cell numbers $N$ with $L/W = 3$. The shape index is $\chi = 3.5$ in simulations.